\documentclass[11pt,a4paper]{article}
\usepackage{jheppub}
\usepackage{pdflscape}
\usepackage{amsmath}
\usepackage{amssymb}
\usepackage{dcolumn}
\usepackage{bm}
\usepackage{color}
\usepackage{epsfig}
\usepackage{amsfonts}
\usepackage{graphicx}
\usepackage{subfigure}
\usepackage{dcolumn}
\usepackage{cancel}

\newcommand{\be}{\begin{equation}}
\newcommand{\ee}{\end{equation}}
\newcommand{\bea}{\begin{eqnarray}}
\newcommand{\eea}{\end{eqnarray}}

\def\be{\begin{equation}}
\def\ee{\end{equation}}
\def\bea{\begin{eqnarray}}
\def\eea{\end{eqnarray}}

\begin{document}

\title{Noether symmetry approach in non-minimal derivative coupling gravity}

\author[a]{Muhammadsorfee Dolohtahe,}
\author[b]{Watcharakorn Srikom,}
\author[c,d]{Phongpichit Channuie,}
\author[a]{Narakorn Kaewkhao}

\affiliation[a]{Department of Physics, Faculty of Science, Prince of Songkla University, Hatyai 90112, Thailand}
\affiliation[b]{ Department of Innovative Technology for Renewable Energy, Faculty of Science and Technology, Suratthani Rajabhat University, Surat Thani, 84100, Thailand}
\affiliation[c]{School of Science, Walailak University, Nakhon Si Thammarat, 80160, Thailand} 
\affiliation[d]{College of Graduate Studies, Walailak University, Nakhon Si Thammarat, 80160, Thailand}

\emailAdd{fee\_dekdee@hotmail.com}
\emailAdd{watcharakorn.sri@sru.ac.th }
\emailAdd{channuie@gmail.com}
\emailAdd{naragorn.k@psu.ac.th}

\abstract {In this work, we examine solutions of the system of equations obtained by applying the Noether gauge symmetry (NGS) and its conserved quantity for the standard general relativity (GR) and the non-minimal derivative coupling (NMDC) cosmological model. We discover two salient features of the solutions. The first one is $a(t)\propto t^{1/3}$ for a kinetic-dominant phase which may emerge before inflationary period at very early time. The second one is a new form of scalar field $\phi(t)$ govern by the exponential cosmological solution for GR and NMDC $\phi_{\rm GR}(t)= \sqrt{c_{1}+c_{2}t+c_{3}e^{-\lambda t}} $ and $\phi_{\rm NMDC}(t)= \sqrt{c_{1}+c_{2}e^{-\lambda_{1}t}+c_{3}e^{-\lambda_{2}t}}$, respectively.}

\keywords{Noether Gauge Symmetry, non-minimal derivative coupling}

\maketitle

\section{Introduction}
Astrophysical observations including Type Ia Supernovae \cite{SupernovaCosmologyProject:1998vns,SupernovaSearchTeam:1998fmf},
cosmic microwave background (CMB) radiation \cite{BICEP2:2015xme,WMAP:2010qai,WMAP:2012nax,Planck:2015fie,Planck:2015sxf,BICEP2:2014owc,BICEP2:2015nss}, large scale structure \cite{SDSS:2003eyi}, baryon acoustic oscillations (BAO) \cite{SDSS:2004kqt} as well as weak lensing \cite{SDSS:2005xqv} make a strong evidence that the expansion of the universe is presently accelerating. In spite its successes, the so-called Lambda cold dark matter ($\Lambda$CDM) \cite{Planck:2018nkj} is plagued by the cosmological problem \cite{Weinberg:2000yb} and the coincident problem \cite{Velten:2014nra}. The phase of late-time cosmic acceleration receives lots of attention. However, the introduction of the
so-called “dark energy (DE)” in the context of conventional general relativity is one of promising explanations. Additionally, another possible scenario is to engineer Einstein gravity on the large-scale methodology. There were some reviewed articles published so far regarding the mentioned issues, see for example \cite{Nojiri:2010wj,Nojiri:2006ri,Capozziello:2011et,Bamba:2015uma} and references therein. However, very little is known about the DE sector of the universe and it possesses one of the unsolved
problems in physics.

Alternative paradigms by engineering the Einstein field equations either in the geometric part or in the stress-energy tensor are widely accepted to explain effects of dark ingredients \cite{Copeland:2006wr}. The $f(R)$ theories of gravity deserves as one of the simplest modifications to the standard general relativity. Here the Lagrangian density of $f$ is an arbitrary function of the scalar curvature $R$ \cite{Bergmann:1968ve,Buchdahl:1970ynr}. It is worth noting that there were rigorous reviews on $f(R)$ theories \cite{Sotiriou:2008rp, DeFelice:2010aj} as well as on Born–Infeld inspired modifications of gravity \cite{BeltranJimenez:2017doy}. In ref.\cite{Nojiri:2017ncd}, the authors investigated the cosmological implications of the modified theories of gravity on inflation, bounce and late-time evolution. Apart from these modified theories of gravity, theories of non-minimal derivative coupling to gravity attract much attention of theoretical and phenomenological points of view, see, e.g., \cite{Amendola:1993uh, Capozziello:1999xt, Capozziello:1999uwa, Granda:2010hb, Granda:2010ex,Sushkov:2009hk, Saridakis:2010mf, Skugoreva:2013ooa, Sushkov:2012za, Koutsoumbas:2013boa, Gumjudpai:2015vio, Gumjudpai:2016frh}. More specifically, their applications on inflation and its consequences were proposed by a clump of authors \cite{Granda:2011zk,Germani:2010gm,Sadjadi:2012zp,Tsujikawa:2012mk,Ema:2015oaa,Yang:2015pga,Myung:2015tga, MohseniSadjadi:2013iou}.  

The important role of the Noether symmetry in cosmology has received increasing attention within decades in order to select the viable models \cite{Capozziello:1994dn}. Conserved quantities of the system, as well as unknown functions, can be determined
with the help of the Noether symmetry approach. More specifically, by using the Noether symmetry, we can obtain the exact solutions. The Noether symmetry approach has been applied to study various cosmological scenarios so far including nonlocal $f(T)$ gravity \cite{Channuie:2017txg,Nurbaki:2020dgw,Jamil:2012fs}, viable mimetic $f(R)$ and $f(R,T)$ theories \cite{Momeni:2015gka}, $f(R)$ cosmology \cite{Capozziello:2008ch}, the cosmological alpha-attractors \cite{Kaewkhao:2017evn}, $f(\mathcal {G})$ theory \cite{Bajardi:2020osh}. Moreover, the exact solutions for potential functions, scalar field and the scale factors in the Bianchi
models have been investigated in Refs.\cite{Camci:2007zz,Jamil:2012zm,Channuie:2018now} and the solutions of the field equations of $f(R)$ gravity are investigated in static cylindrically symmetric space-time using the Noether symmetry technique \cite{Oz:2021nmj}.

The second kind of Noether symmetry approach
for cosmological studies in the literature is the
so-called Noether gauge symmetry (NGS) approach \cite{Aslam:2013pga,Hussain:2011wa,Kucuakca:2011np}. It is a more generalization of the conventional one. Very recently, the authors of Ref.\cite{Kanesom:2021ytb} have discussed the NGS
approach for the Eddington-inspired Born–Infeld theory. In the present work, we study a formal framework of the non-minimal derivative coupling (NMDC) gravity scenario through the NGS approach and
present a detailed calculation of the point-like Lagrangian. The point-like Lagrangian of the Einstein-Hilbert action including the non-minimal derivative coupling (NMDC) sector are examined with spatially flat FLRW spacetime in which matters in such a universe only has a scalar field and a matter field. The latter model is expected to quantify to what extend the field kinetic term affects the evolution of the universe.
 
This paper is organized as follows: We will start by making a short recap of a formal framework of NMDC gravity and derive
the point-like Lagrangian for underlying theory in Section \ref{nmdc}. In Section \ref{secHess}, we study a Hessian matrix and quantify the Euler-Lagrange equations and  Hamiltonian equations of the Einstein (GR) and the NMDC universes. In Section \ref{NGS}, the NGS approach for the GR and NMDC is discussed. We discuss exact cosmological solutions of both theories based with the help of the Noether symmetries of point-like Lagrangian. Finally, we conclude our findings in the last section.

\section{Non-minimal-derivative coupling gravity}\label{nmdc}
The non-minimal derivative coupling (NMDC) gravity model is a special case of the fifth term of Horndeski Lagrangian density that is given as follows:
\begin{eqnarray}
\mathcal{L}_{5}=G_{5}(\phi,X)G^{\mu\nu}\partial_{\mu}\partial_{\nu}\phi,
\end{eqnarray}
where $X=-\frac{1}{2}\partial_{\mu}\phi\partial^{\mu}\phi$. If we set $G_{5}(\phi,X)=-\phi/(M^{2})$ where $M$ is the a new energy scale in the theory. Performing an integration by parts, we obtain
\begin{eqnarray}
\mathcal{L}_{5, \rm NMDC}&=& \cancel{ \partial_{\mu}\Big[(\phi,X)G^{\mu\nu}\partial_{\nu}\phi\Big]}-(\partial_{\nu}\phi)\partial_{\mu}G_{5}(\phi)G^{\mu\nu},\\ \nonumber
&=& \frac{1}{M^{2}}G^{\mu\nu}\partial_{\nu}\phi\partial_{\mu}\phi\,.
\end{eqnarray}
The appearance of minus sign in front of $\frac{1}{M^{2}}G^{\mu\nu}$ is due to the avoidance of ghosts in the scalar field sector, i.e., $\dot{\phi}^{2}+2V(\phi) >-\frac{1}{M^{2}}G_{00}g^{00}g^{00}\dot{\phi}^{2}=-3(\frac{\dot{a}}{a})^{2}\dot{\phi}^{2}.$ Hence, the potential term dominates over the NMDC term \cite{Tsujikawa:2012mk}. The NMDC Lagrangian was first proposed in \cite{Amendola:1993uh} and the action is given by
\bea\label{NMDC action}
S_{\rm NMDC}(g)=\int d^4x \sqrt{-g} \bigg[{R}-(\epsilon g_{\mu \nu}+\kappa G_{\mu \nu})\phi^{,\mu}\phi^{,\nu}-2V(\phi) \bigg]+S_{m}(g_{\mu\nu},\Psi),
\eea
where $\kappa\equiv M^{-2}$ is a NMDC free parameter that has dimension of $[M^{-2}_{P}]$, $S_{m}(g_{\mu\nu},\Psi)$ denotes the matter field action. The spatially flat FLRW metric can be written as
\bea\label{metric g}
ds^{2}_{g} &=& g_{\mu\nu}dx^{\mu}dx^{\nu}= -dt^{2}+a(t)^{2}d{\vec{x}}^{2},
\eea
where $\upsilon_{0}\equiv d^{3}x$ is the spatial volume after a proper compactification for spatial flat section. $\epsilon=+1$ and $\epsilon=-1$ denote an ordinary scalar field and phantom scalar field, respectively. 
The point-like Lagrangian can be extracted from Eq.(\ref{NMDC action}) to yield
\begin{equation} \label{Point-like Lagrangian NMDC 1}
\mathcal{L}_{\rm NMDC}(a,\phi,\dot{a},\dot{\phi})=-6a\dot{a}^2+\bigg({{\epsilon}a^3\dot{\phi}^2}-3\kappa a\dot{a}^2
\dot{\phi}^2 \bigg) -2a^3 V(\phi)-2\rho_{m}(a)a^3.
\end{equation}
It is easy to see that the number of configuration space (n) (or the minisuperspace) is equal to 2 because of the appearance of variables $\{a(t),\phi(t)\}$ in $\mathcal{L}_{\rm NMDC}$.
\section{Hessian matrix, EL equations \& NMDC universe }\label{secHess}

The configuration space variables and their time derivative of both gravity models are $q^{i}= \{a,\phi\}$  and $\dot{q}^{i}=\frac{dq_{i}}{dt} =\{\dot{a},\dot{\phi}\}.$   The Hessian matrix for GR  by using the point-like Lagrangian, $\mathcal{L}_{GR}=-6a\dot{a}^{2}+\epsilon a^{3}\dot{\phi}^{2}-2a^{3}V(\phi)-2\rho_{m}(a)a^{3},$ can be expressed as
\begin{eqnarray}
[W_{ij}]_{\rm{GR}}=\left[
                       \begin{array}{cc}
                         \frac{\partial^{2}L}{\partial \dot{a}^{2}} & \frac{\partial^{2}L}{\partial \dot{a}\partial\dot{\phi}} \\
                         \frac{\partial^{2}L}{\partial \dot{\phi}\partial\dot{a}}  &\frac{\partial^{2}L}{\partial \dot{\phi}^{2}}
                       \end{array}
                     \right]=
                    \left[
                       \begin{array}{cc} 
                   -12a & 0 \\  
                  
                    0 & 2\epsilon a^{3} \\
                        \end{array}
                     \right].
\end{eqnarray}
The determinant of the Hessian matrix of the GR Lagrangian is $\rm{det}[W_{ij,GR}] =-24\epsilon a^{4}\neq 0.$ 
and the Hessian matrix for NMDC can be expressed as
\begin{eqnarray}
[W_{ij}]_{\rm NMDC}=\left[
                       \begin{array}{cc}
                         \frac{\partial^{2}{L}}{\partial \dot{a}^{2}} & \frac{\partial^{2}{L}}{\partial \dot{a}\partial \dot{\phi}} \\
                          \frac{\partial^{2}{L}}{\partial \dot{\phi}\partial \dot{a}}  &\frac{\partial^{2}{L}}{\partial \dot{\phi}^{2}}
                       \end{array}
                     \right]=
                    \left[
                       \begin{array}{cc} 
                   -12a-6\kappa a\dot{\phi}^2 &  -6\kappa a\dot{a}\dot{\phi} \\  
                  
                   -6\kappa a\dot{a}\dot{\phi} &  2\epsilon{a^3}-6\kappa a\dot{a}^{2} \\
                        \end{array}
                     \right].
\end{eqnarray}
The determinant of the Hessian matrix of NMDC Lagrangian are $\rm{det}[W_{ij,NMDC}] =-24\epsilon a^{4}+72\kappa a^{2}\dot{a}^{2}-12\kappa\epsilon a^{4}\dot{\phi}^{2}\neq 0.$ Without the contribution of NMDC free parameter ($\kappa$), this can be clearly reduced to the parameters derived in GR case. 
Mathematically, the fact that the determinant of the Hessian matrix is not equal to zero is called a regular or a non-degenerate Lagrangian. 
It is important to note that a key concept of a gauge field theory is the general solution of the equations of motion that contains arbitrary functions of time and the canonical variables are not all independent but relate to each other by the constraint equations \cite{Henneaux:1992ig}.   
The Hamiltonian constraint equation can be straightforwardly derived from the canonical momenta via the Legendre transformation and Lagrangian as follows:
\begin{eqnarray}\label{Hamiltonian constraint GR}
\mathcal{H}_{\rm GR}&=&\frac{\partial{\mathcal{L}_{\rm GR}}}{\partial \dot{q_i}}\dot{q_i}-\mathcal{L}_{\rm GR}
=\frac{\partial \mathcal{L}_{\rm GR}}{\partial \dot{a}}\dot{a}+\frac{\partial \mathcal{L}_{\rm GR}}{\partial \dot{\phi}}\dot{\phi}-{\mathcal{L}_{\rm GR}},\nonumber \\
&=& p_{a}\dot{a}+p_{\phi}\dot{\phi}-{\mathcal{L}_{\rm GR}},\nonumber \\
&=&\Big(-{12a\dot{a}} \Big)\dot{a}-( 3\epsilon \kappa a \dot{a}^2 \dot{\phi})\dot{\phi}+2\epsilon a^{3}\dot{\phi}-{\mathcal{L}_{\rm GR}}=0, \nonumber\\
&=& -{6a\dot{a}^2} +{\epsilon a^3 \dot{\phi}^2 } +2a^3V(\phi) +2\rho_m(a)a^3=0.
\end{eqnarray}
The energy function that is a constant of motions is given by $E_{\mathcal{L}} \equiv\frac{\partial{\mathcal{L}}}{\partial \dot{q}_{i}}\dot{q}_{i}-\mathcal{L}=p_{i}\dot{q}_{i}-{\cal{L}}\equiv {\mathcal{H}}$ where $p_{i}$ is the canonical momenta. This condition can be rearranged to give the  Friedmann equation in GR case as follows:
\begin{eqnarray}
3H_{\rm GR}^{2}=\rho_{m}(a)+\frac{\epsilon}{2}\dot{\phi}^{2}+V(\phi).
\end{eqnarray}
The Hamiltonian constraint equation for NMDC model can be found similarly as follows:
\begin{eqnarray}\label{Hamiltonian constraint NMDC}
\mathcal{H}_{\rm NMDC}&=&\frac{\partial{\mathcal{L}_{\rm NMDC}}}{\partial \dot{q_i}}\dot{q_i}-\mathcal{L}_{\rm NMDC}
=\frac{\partial \mathcal{L}_{\rm NMDC}}{\partial \dot{a}}\dot{a}+\frac{\partial \mathcal{L}_{\rm NMDC}}{\partial \dot{\phi}}\dot{\phi}-{\mathcal{L}_{\rm NMDC}},\nonumber \\
&=& p_{a}\dot{a}+p_{\phi}\dot{\phi}-{\mathcal{L}_{\rm NMDC}},\nonumber \\
&=&\Big(-{12a\dot{a}}-6\epsilon\kappa a\dot{a}\dot{\phi}^2 \Big)\dot{a}-( 3\epsilon \kappa a \dot{a}^2 \dot{\phi})\dot{\phi}+2\epsilon a^{3}\dot{\phi}-{\mathcal{L}_{\rm NMDC}}=0, \nonumber\\
&=& -{6a\dot{a}^2} -{9\epsilon \kappa a \dot{a}^2 \dot{\phi}^2} +{\epsilon a^3 \dot{\phi}^2 } +2a^3V(\phi) +2\rho_m(a)a^3=0.
\end{eqnarray}
This condition can be rearranged to gives the modified Friedmann equation\cite{Avdeev:2021von} as 
\begin{eqnarray}
3H^{2}=\rho_{m}(a)+\frac{\epsilon}{2}\dot{\phi}^{2}+V(\phi)-\frac{9}{2}\epsilon\kappa H^{2}\dot{\phi}^{2}.
\end{eqnarray}
It's easy to notice that it brings up the Fridmann equation  for GR when $\kappa=0$. The total NMDC Hamiltonian $(\mathcal{H}_{\rm NMDC})$ can be used to evaluate an evolution invoking the Hamiltonian equations of motion as follows:
\begin{eqnarray}\label{Hamiltonian eqs EiBI}
p_{a}&=&\frac{\partial \mathcal{L}_{\rm NMDC}}{\partial{\dot{a}}}=\dot{a}\Big[-12a-6\epsilon \kappa a\dot{\phi}^{2}  \Big],\\
p_{\phi}&=&\frac{\partial \mathcal{L}_{\rm NMDC}}{\partial{\dot{\phi}}}=\dot{\phi}\Big[ 2\epsilon a^{3}-6\epsilon\kappa a \dot{a}^{2} \Big]\\
\dot{a}&=&\frac{\partial \mathcal{H}_{\rm NMDC}}{\partial p_{a}}=\frac{-p_{a}}{\Big[12a+6\epsilon \kappa a\dot{\phi}^{2}\Big]}\\
\dot{\phi}&=&\frac{\partial \mathcal{H}_{\rm NMDC}}{\partial p_{\phi}}=\frac{p_{\phi}}{\Big[ 2\epsilon a^{3}-6\epsilon\kappa a\dot{a}^{2}\Big]}\\
\dot{p}_{a}&=&-\frac{\partial \mathcal{H}_{\rm NMDC}}{\partial a}=6a^{2}+9\epsilon \kappa \dot{a}^{2}\dot{\phi}^{2}-3\epsilon a^{2}\dot{\phi}^{2}-6a^{2}V(\phi)-2\rho'_{m}(a)a^{3}-6\rho_{m}(a)a^{2}\\
\dot{p}_{\phi}&=&-\frac{\partial \mathcal{H}_{\rm NMDC}}{\partial \phi}=-2a^{3}V'(\phi).
\end{eqnarray}
In order to obtain the dynamic solutions, we have to calculate the Euler-Lagrange equations for $a(t)$ and $\phi(t)$ which gives the same results as Eq.(2)-(4) of Ref \cite{Avdeev:2021von} as shown below:
\begin{eqnarray}\label{ELEiBI}
&&6\dot{a}^{2}-3\epsilon a^{2}\dot{\phi}^{2}+3\kappa\dot{a}^{2}{\dot{\phi}^{2}}+6a^{2}V(\phi)+2\rho'_{m}(a)a^{3}+6\rho_{m}(a)a^{2}-12\dot{a}^{2}-12a\ddot{a}-6\kappa\dot{a}^{2}\dot{\phi}^{2}\nonumber\\ &&-6\kappa a\ddot{a}\dot{\phi}^{2}- 12\kappa a\dot{a}\dot{\phi}\ddot{\phi}=0.
\end{eqnarray}
Rearranging, this shows that
\begin{eqnarray}\label{ELEiBI}
3H^{2}+2\dot{H}&=&-\frac{\dot{\phi}^{2}}{2}
\Big[\epsilon+\kappa(2\dot{H}+3H^{2}+4H\frac{\ddot{\phi}}{\dot{\phi}})  \Big]
+V(\phi)+\frac{\rho'_{m}(a)a}{3}+\rho_{m},\nonumber \\
&=&-\frac{\dot{\phi}^{2}}{2}
\Big[\epsilon+\kappa(2\dot{H}+3H^{2}+4H\frac{\ddot{\phi}}{\dot{\phi}})  \Big]
+V(\phi)-p_{m},
\end{eqnarray}
where the fluid equation \cite{Bouhmadi-Lopez:2016dcf,{Babara Ryden 2017}}, $\rho'(a)\equiv d\rho_{m}/da=-3(\rho_{m}+p_{m})/a,$ has been used to get the last line of Eq.(\ref{ELEiBI}).

The modified Klien-Gordon equation for NMDC is directly calculated from the Euler-Lagrange equation for scalar field. Thus this gives 
\begin{eqnarray}
2a^{3}V_{\phi}+6\epsilon a^{2}\dot{a}\dot{\phi}+2\epsilon a^{3}\ddot{\phi}-6\kappa \dot{a}^{3}\dot{\phi}-12\kappa a\dot{a}\ddot{a}\dot{\phi}-6\kappa a\dot{a}^{2}\ddot{\phi}=0.
\end{eqnarray}
Having used the relation $\frac{\ddot{a}}{a}=\dot{H}+H^{2}$, it can be shown that
\begin{eqnarray}
\epsilon(\ddot{\phi}+3H\dot{\phi})-3\kappa\Big( H^{2}\ddot{\phi}+2H\dot{H}\dot{\phi}+3H^{3}\dot{\phi} \Big)=-V_{\phi}.
\end{eqnarray}

\section{Noether gauge symmetries}
\label{NGS}
In this section, we employ the Noether gauge symmetries to figure out exact solutions of the systems both in the standard GR cosmology and in the NMDC universe.
\subsection{Standard GR cosmology}
The Lagrangian for GR is written as
\begin{equation}
\mathcal{L}_{GR}=\sqrt{-g}{R}+\mathcal{L}_{\phi}+\mathcal{L}_{m},
\end{equation}
where the second term is the scalar field Lagrangian and the last term is the matter field one. We can derive the point-like Lagrangian of GR to obtain
\begin{equation}
\mathcal{L}_{GR}= {-6a\dot{a}^2}+\epsilon a^{3}\dot{\phi}^{2}-2a^{3}V(\phi)-2a^{3}\rho_{m}.
\end{equation}
The approach of Noether gauge symmetry can be apploied to Eq.(\ref{Point-like Lagrangian NMDC 1}) aiming to specify cosmological functions of the standard GR and NMDC gravity. A vector field in this approach can be written as
\begin{equation}\label{vector field NGS}
\mathrm{X}_{\rm NGS}=\tau\frac{\partial}{\partial t}+\alpha\frac{\partial}{\partial a}+\varphi\frac{\partial}{\partial \phi}.
\end{equation}
The first prolongation of $N_{\rm NGS}$ reads
\begin{eqnarray}\label{first prolongation}
 \mathrm{X}^{[1]}_{\rm NGS}&=&\mathrm{X}_{\rm NGS}+\dot{\alpha}\frac{\partial}{\partial \dot{a}}+\dot{\varphi}\frac{\partial}{\partial \dot{\phi}},
\end{eqnarray}
where the undetermined  parameter $\tau$  is a function of  $\{t,a,\phi\}$. The time derivative for  $\alpha(t,a,\phi)$ and $\varphi(t,a,\phi)$ are defined as
\begin{eqnarray}\label{dot alphasa}
\dot{\alpha}(t,a,\phi)&=&\mathrm{D}_{t}\alpha-\dot{a}\mathrm{D}_{t}\tau,\nonumber \\
\dot{\varphi}(t,\phi)&=&\mathrm{D}_{t}\varphi-\dot{\phi}\mathrm{D}_{t}\tau.
\end{eqnarray}
Here
$\mathrm{D}_t$ is the operator of a total differentiation with respect to $t$, i.e.
\begin{equation}
    \mathrm{D}_t =\frac{\partial}{\partial t}+\dot{a}\frac{\partial}{\partial a} +\dot{\phi}\frac{\partial}{\partial \phi}.
\end{equation}
The vector field $\mathrm{X}_{\rm NGS}$  is a NGS of a Lagrangian ${\mathcal{L}}(t,a,\phi,\dot{a},\dot{\phi})$, if there exists a gauge function $\mathrm{B}(t, a, \phi)$ which obeys the Rund-Trautmann identity, see \cite{SARLET-1981,Leone R-2015,Mukherjee:2021bna} for explicit derivation
\begin{eqnarray}\label{NGS gaugcd}
    \mathrm{X}^{[1]}_{NGS}{\mathcal{L}}+{\mathcal{L}}\mathrm{D}_t\tau=\mathrm{D}_t \mathrm{B}.
\end{eqnarray}
For NSG without gauge term, i.e. $\mathrm{B}=0$, it requires that $\tau=0$. Therefore Eq.(\ref{NGS gaugcd}) is reduced to $\textit{\pounds}_{\mathrm{X}^{[1]}_{\rm NGS}}{\mathcal{L}}=0$ that is the condition for Noether symmetry \cite{Kucuakca:2011np}.\\
The Noether gauge condition yields
\begin{equation}
X^{[1]}_{NGS}\mathcal{L}_{GR}+\mathcal{L}_{GR} D_{t}\tau=D_{t}B
\end{equation}
Using the Noether gauge condition to $ \mathcal{L}_{GR} $, we have 24 terms which can be expressed as follows:
\begin{eqnarray}
D_tB &=&-6a^2V(\phi)\alpha -6a^2\alpha\rho_m(a)  -{6\alpha \dot{a}^2}
  -2a^3\varphi V'(\phi) +{3\epsilon a^2\alpha\dot{\phi}^2}  -2a^3\alpha\rho'_m(a) \nonumber\\  &&
 -{12a\dot{a}\alpha_t} -2a^3V(\phi)\tau_t -2a^3\rho_m(a)\tau_t +{6a\dot{a}^2\tau_t} -{\epsilon a^3\dot{\phi}^2\tau_t}
 +{2\epsilon a^3\dot{\phi}\varphi_t} -{12a\dot{a}\dot{\phi}\alpha_\phi}\nonumber \\ && -2a^3V(\phi)\dot{\phi}\tau_\phi -2a^3\rho_m(a)\dot{\phi}\tau_\phi +{6a\dot{a}^2\dot{\phi}\tau_\phi} -{\epsilon a^3\dot{\phi}^3\tau_\phi} +{2\epsilon a^3\dot{\phi}^2\varphi_\phi} -12a\dot{a}^{2}\alpha_{a}\nonumber\\  &&-2a^{3}V(\phi)\dot{a}\tau_{a}-2a^{3}\rho_{m}(a)\dot{a}\tau_{a}+6a\dot{a}^{3}\tau_{a}-\epsilon a^{3}\dot{a}\dot{\phi}^{2}\tau_{a}+2\epsilon a^{3}\dot{a}\dot{\phi}\varphi_{a}
\end{eqnarray}
 
After separation of monomials and polynomials in term of $ \dot{a}^2, \dot{\phi}^2, \dot{a}, \dot{\phi}, \dot{a}\dot{\phi}, \dot{a}^2\dot{\phi},\dot{\phi}^3, \dot{a}^3$ and $\dot{a}\dot{\phi}^2$, the constraints equations and the PDEs can be given as
\begin{eqnarray}\label{alpha a1}
\varphi_a=\alpha_\phi=\tau_\phi=\tau_a=0
\end{eqnarray}
\begin{eqnarray}\label{alpha a2}
\   \alpha+2a\alpha_a-a\tau_t=0
\end{eqnarray}
\begin{equation}
    3\alpha+2a\varphi_\phi-a\tau_t=0
\end{equation}
\begin{eqnarray}
6\alpha_{\phi}=\epsilon a^{2}\varphi_{a}=0
\end{eqnarray}
\begin{equation}\label{Ba}
B_a=-12a\alpha_{t}
\end{equation}
\begin{equation}\label{Bphi}
B_\phi=2\epsilon a^3\varphi_t
\end{equation}
\begin{eqnarray}\label{Bt_gaugeterm}
B_t&=&-\alpha\big[ 6a^2V(\phi)+6a^2\rho_m(a)  +2a^3\rho '_m(a)  \Big] -2\varphi a^3 V'(\phi)\nonumber \\ &&
-\tau_t \Big[ 2a^3V(\phi)+2a^3\rho_m(a) \Big] .
\end{eqnarray}

\begin{table}[h]
 \begin{tabular}{||c c c c ||} 
 \hline
$\alpha(a)$ & $\varphi(\phi)$ & $\tau(t)$ & $\rm{add.\,\, conditions}$\\ 
\hline\hline
 $ c_{1}a$ & $ c_{2}\phi$ & $3c_{1}t+c_{3}$& NO  \\ 
 \hline
 $c_{1}a$ & $e^{c_{2}\phi}$ &  $c_{2}t^{c_{4}}+3c_{1}t+c_{5}$ & $\phi(t)=c_{3}\ln t,\,\, c_{4}\equiv c_{2}c_{3}$ \\
 \hline
 $c_{1}\ln a$ & $c_{2 \ln \phi}$ &  $\int{\big( 2c_{1}\frac{\ln a}{a}+\frac{c_{1}}{a} +\frac{c_{2}}{\phi}\big) dt }$ & $c_{2}=\frac{\phi}{a}(\ln a-c_{1})$ \\
 \hline
\end{tabular}
   \caption{We show possible parameters of the system $\alpha,\,\varphi$ and $\tau$ and their relations.}
    \label{Table}
    \end{table}
Because there has no additional conditions, we are interest in examining only for a linear form of solutions. Using the fact that $\alpha_{a}= c_{1}$ and
$\varphi_{\phi} = c_{2}$,  this gives $\tau_{t}=3c_{1}$.
The boundary term partly derived from Eq.(\ref{Ba}) and Eq.(\ref{Bphi}) is expressed as
\begin{eqnarray}\label{BaBphi partly}
B_{(a,\phi)}=-6c_{1}a^{2}\dot{a}+2c_{2}\epsilon a^{3}\phi \dot{\phi}.
\end{eqnarray}
The constant of motion for GR case is given by
\begin{eqnarray}
\Sigma_{0,\rm GR}&=&\alpha\frac{\partial \mathcal{L}}{\partial \dot{a}}+\varphi\frac{\partial \mathcal{L}}{\partial \dot{\phi}},\nonumber \\
&=&\alpha( -12a\dot{a} )+\varphi(2\epsilon a^{3}\dot{\phi}).\nonumber\\
&=& -12c_{1}a^{2}\dot{a}+2c_{2}\epsilon a^{3}\phi\dot{\phi}\,. \label{constant MGR}
\end{eqnarray}
A first integral or a Noether integral of the system  or a conserved quantity associated with $X$ can be derived using
\begin{eqnarray}\label{Noether integral}
I(t,q^{i},\dot{q}^{i})=-\tau(t,q^{i})\Big( \dot{q}^{i}\frac{\partial \mathcal{L}}{\partial \dot{q}^{i}}-\mathcal{L}\Big)+\eta^{i}\frac{\partial \mathcal{L}}{\partial \dot{q}^{i}}-B(t,q^{i})\,.
\end{eqnarray}
It is easy to see that when setting $\tau=0$ or $\Big( \dot{q}^{i}\frac{\partial \mathcal{L}}{\partial \dot{q}^{i}}-\mathcal{L}\Big)=0$
\begin{eqnarray}\label{first integral2}
I(t,q^{i},\dot{q}^{i})&=&\eta^{i}\frac{\partial \mathcal{L}}{\partial \dot{q}^{i}}-B(t,q^{i}),\nonumber \\
&=& \Sigma_{0}-B(t,q^{i}).
\end{eqnarray}
where in GR case $\tau(t)=3c_{1}t+c_{3}$ and $\eta^{i}=\{\alpha(a)=c_{1}a,\varphi(\phi)=c_{2}\phi\}.$
The Lagrangian-related Noether vector X for GR is
\begin{eqnarray}
X=\xi(t,q^{k})\frac{\partial \mathcal{L}}{\partial t}+\eta^{i}(t,q^{k})\frac{\partial}{\partial q^{i}}.
\end{eqnarray}
This gives
\begin{eqnarray}
X_{\rm GR}=(3c_{1}t+c_{3})\partial_{t}+c_{1}a\partial_{a}+c_{2}\phi\partial_{\phi}.
\end{eqnarray}
The corresponding generators of a Noether point symmetry  are
\begin{eqnarray}
X_{1}&=&\partial_{t},\nonumber \\
X_{2}&=&3t\partial_{t}+a\partial_{a}+\phi\partial_{\phi}.
\end{eqnarray}
The simple commutative algebra of the two symmetry generators is
\begin{eqnarray}
[X_{1},X_{2}]=0.
\end{eqnarray}
 Since it admits the Noether symmetry $\partial_{t}$ ,the first corresponding conserved quantity ($I_{1}$) related to the energy conservation via the time translation symmetry is
\begin{eqnarray}
I_{1,\rm GR}=E_{{\cal L}_{\rm GR}}=\mathcal{H}_{\rm GR}=0.
\end{eqnarray}
The second Noether integral is
\begin{eqnarray}
I_{2, \rm GR}&=& -\tau(t,a,\phi)\Big[ \dot{a}\frac{\partial \mathcal{L}}{\partial{\dot{a}}}+\dot{\phi}\frac{\partial \mathcal{L}}{\partial{\dot{\phi}}}-\mathcal{L} \Big]+\Sigma_{0, GR}-B(t,a,\phi),\nonumber \\
&=&-(3c_{1}t+c_{3})\Big[ -12a\dot{a}^{2}+2\epsilon a^{3}\dot{\phi}^{2}+6a\dot{a}^{2}-\epsilon a^{3}\dot{\phi}^{2}+2a^{3}V(\phi)+2a^{3}\rho_{m} \Big]\nonumber \\
&& -12c_{1}a^{2}\dot{a}+\cancel{2c_{2}\epsilon a^{3}\phi\dot{\phi}}+6c_{1}a^{2}\dot{a}\cancel{-2c_{2}\epsilon a^{3}\phi\dot{\phi}},\nonumber \\
&=&-(3c_{1}t+c_{3})\Big[ -6a\dot{a}^{2} +\epsilon a^{3}\dot{\phi}^{2}+2a^{3}V(\phi)+2a^{3}\rho_{m})\Big]-6c_{1}a^{2}\dot{a}.
\end{eqnarray}
Using Eq.(\ref{Hamiltonian constraint GR}), i.e. $\mathcal{H}_{\rm GR}=0$, this gives
\begin{eqnarray}
I_{2,\rm GR}&=&(3c_{1}t+c_{3}){\mathcal{H}_{\rm GR}}-6c_{1}a^{2}\dot{a},\nonumber \\
&=&-6c_{1}a^{2}\dot{a}.
\end{eqnarray}
That leads to 
\begin{eqnarray}
a(t)&=&\Big(-\frac{I_{2,\rm GR}}{2c_{2}}\Big)^{1/3}t^{1/3},\nonumber \\
a(t)&=&a_{0}t^{1/3}.
\end{eqnarray}
Comparing to scale factor for single-component universe see section 5.3 of Ref. \cite{Babara Ryden 2017},
\begin{eqnarray}
a(t)=(\frac{t}{t_{0}})^{\frac{2}{3+3w}},
\end{eqnarray}
 this yields the equation of state parameter for dominance role of kinetic part of the scalar field rather than it  potential, i.e. $\dot{\phi}^{2}\gg V(\phi)$
\begin{eqnarray}
w=1,
\end{eqnarray}
this indicate the existence a kinetic dominance (KD) phase  after the big bang but before the slow roll condition took place. The form of scalar field that corresponds to that scale factor is $\phi \propto t^{1/6}$ \cite{Handley:2014bqa}\\
For $\alpha(a)=c_{1}\ln a$ and $\varphi(\phi)=c_{2}\ln \phi$, this gives $\alpha_{\phi}=0$, $\varphi_{a}=\frac{\partial}{\partial a}(c_{2}\ln \phi)=0$
and $c_{2} = \frac{\phi}{a}(\ln a-c_{1})$ as shown as an example in Table (\ref{Table}).
The  constant of motion and gauge function related to this form of solutions take the form
\begin{eqnarray}
\Sigma_{0,GR(2)}&=&-12c_{1}a\dot{a}\ln a+2\epsilon c_{2}a^{3}\dot{\phi}\ln {\phi},\nonumber \\
B_{(a,\phi)}&=&-12c_{1}a\dot{a}+2\epsilon c_{2}a^{3}\dot{\phi}\ln \phi\,,
\end{eqnarray}
respectively. From Eq.(\ref{first integral2}), we find
\begin{eqnarray}
\frac{I_{2,GR(2)}}{12c_{1}}t=c_{3}=\frac{3a^{2}}{4}-\frac{1}{2}
a^{2}\ln a
\end{eqnarray}
which leads to a scale factor written of the form
\begin{eqnarray}
a(t)= \frac{\pm2\sqrt{|c_{3}|t}}{\sqrt{\text{production} \ln(\frac{4c_{3}t}{e^{3}})}}\,.
\end{eqnarray}
Here we will not elaborate on this case further. 
\subsection{NMDC universe}
For the NMDC universe, the Lagrangian for NMDC gravity is written as
\begin{equation}
    \mathcal{L}_{NMDC}=\sqrt{-g}\Big[ {R}-(\epsilon g_{\mu\nu}+\kappa G_{\mu\nu})\nabla^{\mu}\phi \nabla^{\nu}\phi -2V(\phi)\Big],
\end{equation}
in which it can illustrate the point-like Lagrangian of NMDC as
\begin{equation}
\mathcal{L}_{NMDC}={-6a\dot{a}^2}+\bigg({\epsilon}{a^3\dot{\phi}^2}-3\kappa a\dot{a}^2
\dot{\phi}^2 \bigg) -2a^3 V(\phi)-2a^3\rho_{m}(a).
\end{equation}

The Noether gauge condition in this case yields 
\begin{equation}
X^{[1]}_{NGS}\mathcal{L}_{NMDC}+\mathcal{L}_{NMDC} D_{t}\tau=D_{t}B
\end{equation}
Using the Noether gauge condition to $ \mathcal{L}_{NMDC} $, this gives 34 terms which can be expressed as follows:
\begin{eqnarray}\label{NGS NMDC1}
D_{t}B&=& -6a^{2}V(\phi)\alpha-6a^{2}\alpha\rho_{m}(a)-6\alpha\dot{a}^{2}-2a^{3}\varphi V'(\phi)+3\epsilon a^{2}\alpha\dot{\phi}^{2}-3\kappa\alpha\dot{a}^{2}\dot{\phi}^{2}-2a^{3}\alpha\rho'_{m}\nonumber\\&&-12a\dot{a}\alpha_{t}-6\kappa a\dot{a}\dot{\phi}^{2}\alpha_{t}-2a^{3}V(\phi)\tau_{t}-2a^{3}\rho_{m}(a)\tau_{t}
+6a\dot{a}^{2}\tau_{t}-\epsilon a^{3}\dot{\phi}^{2}\tau_{t}+9\kappa a \dot{a}^{2}\dot{\phi}^{2}\tau_{t}\nonumber \\&&+2\epsilon a^{3}\dot{\phi}\varphi_{t}-6\kappa a\dot{a}^{2}\dot{\phi}\varphi_{t}-12a\dot{a}\dot{\phi}\alpha_{\phi}-6\kappa a\dot{a}\dot{\phi}^{3}\alpha_{\phi}-2a^{3}V(\phi)\dot{\phi}\tau_{\phi}-2a^{3}\rho_{m}(a)\dot{\phi}\tau_{\phi}\\&& +6a\dot{a}^{2}\dot{\phi}\tau_{\phi}-\epsilon a^{3}\dot{\phi}^{3}\tau_{\phi}+9\kappa a\dot{a}^{2}\dot{\phi}^{3}\tau_{\phi}+2\epsilon a^{3}\dot{\phi}^{2}\varphi_{\phi}-6\kappa a \dot{a}^{2}\dot{\phi}^{2}\varphi_{\phi}-12a\dot{a}^{2}\alpha_{a}-6\kappa a\dot{a}^{2}\dot{\phi}^{2}\alpha_{a}\nonumber\\&&-2a^{3}V(\phi)\dot{a}\tau_{a}-2a^{3}\rho_{m}(a)\dot{a}\tau_{a}+
6a\dot{a}^{3}\tau_{a}-\epsilon a^{3}\dot{a}\dot{\phi}^{2}\tau_{a}+9\kappa a\dot{a}^{3}\dot{\phi}^{2}\tau_{a}+2\epsilon a^{3}\dot{a}\dot{\phi}\varphi_{a}-6\kappa a\dot{a}^{3}\dot{\phi}\varphi_{a}.\nonumber
\end{eqnarray}
After separation of polynomials and monomials, we can express each term in a more compact form as follows:
\begin{eqnarray}
\dot{a}^{2}\Big[ -6\alpha+6a\tau_{t}-12a\alpha_{a} \Big]&=&0, \\
\dot{\phi}^{2}\Big[ 3\epsilon a^{2}\alpha-\epsilon a^{3}\tau_{t}+2\epsilon a^{3}\varphi_{\phi}\Big]&=&0,\\
\dot{a}^{2}\dot{\phi}^{2}\Big[ -3\kappa\alpha+9\kappa a\tau_{t}-6\kappa a\varphi_{\phi}-6\kappa a\alpha_{a} \Big]&=&0,\\
\dot{a}\dot{\phi}^{2}\Big[ -6\kappa a\alpha_{t}-\epsilon a^{3}\tau_{a}  \Big]&=&0,\\
\dot{a}^{2}\dot{\phi}\Big[ -6\kappa a \varphi_{t}+6a\tau_{t}   \Big]&=& 0,\\
\dot{a}\dot{\phi}\Big[  -12a\alpha_{\phi}+2\epsilon a^{3}\varphi_{a}\Big]&=& 0,\\
\dot{a}\dot{\phi}^{3}\Big [ -6\kappa a \alpha_{\phi} \Big]&=& 0, \\
 \dot{\phi}^{3}\Big [ -\epsilon a^{3}\tau_{\phi} \Big]&=& 0, \\
\dot{a}^{2}\dot{\phi}^{3}\Big[  9\kappa a \tau_{\phi} \Big]&=&0,\\
\dot{a}^{3}\Big[ 6a\tau_{a}  \Big]&=&0, \\
\dot{a}^{3}\dot{\phi}^{2} \Big[ 9\kappa a \tau_{a} \Big]&=&0,\\
\dot{a}^{3}\dot{\phi}\Big[  -6\kappa a \varphi_{a} \Big]&=& 0,
\end{eqnarray}
\begin{eqnarray}
\dot{a}B_{a}&=&\dot{a}\Big[ -12a\alpha_{t}-2a^{3}V(\phi)\tau_{a}-2a^{3}\rho_{m}(a)\tau_{a} \Big],\\
\dot{\phi}B_{\phi}&=& \dot{\phi}\Big [2\epsilon a^{3}\varphi_{t}-2a^{3}V(\phi)\tau_{\phi}-2a^{3}\rho_{m}(a)\tau_{\phi} \Big], \\
B_{t}&=&-\alpha\Big[ 6a^{2}V(\phi)+6a^{2}\rho_{m}(a)+2a^{3}\rho'_{m}(a)\big]+2\varphi a^{3}V'(\phi)\nonumber\\&+&\tau_{t}\big[ 2a^{3}V(\phi)+2a^{3}\rho'_{m}(a)  \Big].
\end{eqnarray}
The constraint equations and the PDEs can be given as follows:
\begin{equation}\label{cs0}
    \varphi_a=\varphi_t=\alpha_\phi=\alpha_t=\tau_a=\tau_{\phi}=0.
\end{equation}

\begin{equation}\label{cs1}
    \alpha+2a\alpha_a-a\tau_t=0
\end{equation}

\begin{equation}\label{cs2}
    3\alpha+2a\varphi_\phi-a\tau_t=0
\end{equation}

\begin{equation}\label{cs3}
    \alpha+2a\varphi_\phi+2a\alpha_a-3a\tau_{t}=0
\end{equation}

\begin{equation}\label{BaBphizero}
    B_{a}=B_{\phi}=0
\end{equation}
\begin{eqnarray}\label{Bt term}
B_t&=&-\alpha\big[ 6a^2V(\phi)+6a^2\rho_m(a)  +2a^3\rho '_m(a)  \Big] -\varphi \Big[(2a^3 V'(\phi)\Big]\nonumber \\ &&
-\tau_t \Big[ 2a^3V(\phi)+2a^3\rho_m(a) \Big] .
\end{eqnarray}
What is different from GR is the non-vanishing of two terms contributing to NMDC, i.e. $B_{a}=B_{\phi}=0.$ Offsetting this, Eq.(\ref{cs3}) is modified due to the existence of the NMDC term, i.e., 
\begin{eqnarray}
\kappa \dot{a}^{2}\dot{\phi}^{2}\Big( 3\alpha+6a\varphi_{\phi}+6a\alpha_{a}-9a\tau_{t}\Big)=0.
\end{eqnarray} 
Substituting Eq.(\ref{cs1}) into Eq.(\ref{cs3}), this yields the relation $\varphi_{\phi}=\tau_{t}.$ From Eq.(\ref{cs1}) and Eq.(\ref{cs2}), this gives
\begin{eqnarray}
a(t)=-\frac{3\alpha}{\varphi_{\phi}}=-\frac{3\alpha}{3(2\alpha_{a}-\tau_{t})}.
\end{eqnarray}
It should be noted that $\varphi_\phi\neq 0$ and $2\alpha_{a}-\tau_{t}\neq 0.$
Hence we have one more relation given by
\begin{eqnarray}
\varphi_{\phi}=3(2\alpha_{a}-\tau_{t}).
\end{eqnarray}
Using the fact that $\varphi_{\phi}=\tau_{t}$, this leads to the relation of $\tau_{t}$ , $\alpha_{a}$ and $\varphi_{\phi}$ as follows:
\begin{eqnarray}\label{equal solutions}
\alpha_{a}&=&\frac{2}{3}\tau_{t}=\frac{2}{3}\varphi_{\phi},\\
c_{1}&=&\frac{2}{3}c_{2}\,.
\end{eqnarray}
We find that the relation does not allow for the exponential and Logarithmic forms.
Therefore, we assume 
\begin{eqnarray}
\alpha(a)= c_{1}a, \\
\varphi(\phi)=c_{2}\phi,
\end{eqnarray}
this leads to
\begin{eqnarray}
\tau(t)&=&c_{2}t+c_{3},\nonumber\\
\tau_{t}&=&c_{2}.
\end{eqnarray}
The constant of motion for NMDC case reads
\begin{eqnarray}
\Sigma_{0,\rm NMDC}&=&\alpha\frac{\partial \mathcal{L}}{\partial {\dot{a}}}+\varphi\frac{\partial \mathcal{L}}{\partial \dot{\phi}},\\
&=&\alpha\Big[ -12a\dot{a}-6\kappa a\dot{a}\phi^{2} \Big]+\varphi\Big[2\epsilon a^{3}\dot{\phi}-6\kappa a \dot{a}^{2}\phi\Big].\nonumber\\
&=& -12c_{1}a^{2}\dot{a}+2c_{2}\epsilon a^{3}\phi\dot{\phi}-\kappa\Big[6c_{1} a^{2}{\phi}^{2}\dot{a}+6c_{2} a{\phi}^{2}\dot{a}^{2}\Big].
\end{eqnarray}
A first integral of the system  or a conserved quantity associated with $X_{NGS}$ can be derived from Eq.(\ref{Noether integral})
where in NMDC case we have used $\xi=\tau(t)= c_{2}t+c_{3}$ and $\eta^{i}=\{\alpha(a)=c_{1}a,\varphi(\phi)=c_{2}\phi\}.$
The Lagrangian-related Noether vector X for NMDC is
\begin{eqnarray}
X_{\rm NMDC}=\Big( c_{2}t+c_{3}\Big)\partial_{t}+c_{1}a\partial_{a}+c_{2}\phi\partial_{\phi}.
\end{eqnarray}
The corresponding Noether symmetries are
\begin{eqnarray}
X_{1}&=&\partial_{t},\nonumber \\
X_{2}&=&t\partial_{t}+a\partial_{a}+\phi\partial_{\phi}.
\end{eqnarray}
The simple commutative algebra of the two symmetry generators is
\begin{eqnarray}
[X_{1},X_{2}]=0.
\end{eqnarray}
\begin{eqnarray}
I_{1,\rm NMDC}&=&E_{{\cal L}_{\rm NMDC}}=\mathcal{H}_{\rm NMDC}=0,\nonumber \\
I_{2, \rm NMDC}&=& \Big( c_{2}t+c_{3}\Big){\mathcal{H}_{\rm NMDC}}+\Sigma_{\rm NMDC}-B_{\rm NMDC}\nonumber \\
&=& -12c_{1}a^{2}\dot{a}+2c_{2}\epsilon a^{3}\phi\dot{\phi}-\kappa\Big[ 6c_{1}a^{2}\phi^{2}\dot{a}+6c_{2}a\phi^{2}\dot{a}^{2} \Big].\label{interplay a phi}
\end{eqnarray}
Eq.(\ref{interplay a phi}) may shed some light on the interplay between $\phi(t)$ and $a(t).$ From the fact that $\frac{d \Sigma_{0,\rm NMDC}}{dt}=\frac{d I_{2, \rm NMDC}}{dt}=0$ and
the exponential cosmological solution expressed in term of 
\begin{eqnarray}
a(t)&=&e^{\frac{1}{2}H_{0}t},\\
\dot{a}(t)&=&\frac{1}{2}H_{0}e^{\frac{1}{2}H_{0}t},
\end{eqnarray}
where  $H_{0}$ is a constant \cite{Harko:2016xip,Darabi:2013caa}, this
leads to an interesting arrangement shown below:
\begin{eqnarray}\label{ INM eq}
\frac{I_{2,\rm NMDC}}{3c_{1}}e^{-\frac{3}{2}H_{0}t}+\frac{2H_{0}}{c_{1}}-
\epsilon \phi \frac{d\phi}{dt}-\kappa\phi^{2}(H_{0}+\frac{3}{4}H^{2}_{0})=0.
\end{eqnarray}
The positive solution is
\begin{eqnarray}\label{solut I2}
\phi(t)= \frac{\sqrt{ c_{4}+c_{5}e^{-\lambda_{1}t}+c_{6}e^{{-\lambda_{2}t}}}}{\sqrt{c_{7}}}\,.
\end{eqnarray}
For simplicity, we have set
\begin{eqnarray}
c_{4} &\equiv&  -\frac{4\kappa H_{0} }{c_{1}}  (H_{0}+\frac{3}{4}H_{0}^{2})-\frac{2\epsilon \lambda_{1}H_{0}}{c_{1}},\nonumber\\
c_{5}&\equiv& \frac{2\kappa I_{2,\rm NMDC}}{3c_{1}}(H_{0}+\frac{3}{4}H_{0}^{2}),\nonumber\\
c_{6}&\equiv& \kappa\epsilon \lambda_{1} C[1](H_{0}+\frac{3}{4}H^{2}_{0})+2\kappa^{2}C[1](H_{0}+\frac{3}{4}H^{2}_{0})^{2},\\
c_{7}&\equiv& \kappa\Big(H_{0}+\frac{3}{4}H_{0}^{2})(\epsilon \lambda_{1}-2(H_{0}+\frac{3}{4}H_{0}^{2})\Big),\nonumber\\
\lambda_{1}  &\equiv& \frac{3H_{0}}{2},\nonumber\\
\lambda_{2} &\equiv& \frac{2\kappa (H_{0}+\frac{3}{4}H_{0}^{2})}{\epsilon},\nonumber
\end{eqnarray}
where $C[1]$ is arbitrarily constant. 
We finally consider the case of  GR cosmology by applying the same exponential cosmological solution to Eq.(\ref{constant MGR}) and this gives
\begin{eqnarray}\label{constant MGR}
\frac{\Sigma_{0,\rm GR}}{2}e^{-\frac{3}{2}H_{0}t}-3c_{1}H_{0}-c_{2}\epsilon
\phi\frac{ d\phi}{dt}=0.
\end{eqnarray}
The positive solution of Eq.(\ref{constant MGR}) can be simplified into
\begin{eqnarray}
\phi(t)=\sqrt{c_{1}+c_{2}t+c_{3}e^{-\lambda t}},
\end{eqnarray}
where $c_{1}=$ arbitrarily constant, $c_{2}=\frac{6c_{1}H_{0}}{c_{2}\epsilon}$ and $c_{3}=-\frac{2\Sigma_{0,\rm GR}H_{0}}{3\epsilon c_{2}}.$
\section{Conclusion  \label{summary}}
In the present work, we have considered a formal framework of NMDC gravity and derived the point-like Lagrangian for underlying theory. We have studied a Hessian matrix and quantified the Euler-Lagrange equations and  Hamiltonian equations of the Einstein (GR) and the NMDC universes. We further discussed the NGS approach for the GR and NMDC universe. We have studied exact cosmological solutions of both theories with the help of the Noether symmetries of point-like Lagrangian.

 We have assumed the linear forms of $\alpha(a),\varphi(\phi)$ and $\tau(t)$ that are compatible with the structure of the standard GR and explained the possible emergence of the kinetic field dominated phase which might take place before the inflationary stage. We discovered that the NGS condition under the structure of NMDC gravity can eliminate the dependence of the variables $a(t)$ and $\phi(t)$ on a gauge function. Consequently, a gauge function is at most dependent on the time variable only. The simplest form of $B(t)$ can be written as $B(t)=c_{8}t+c_{9}$. Interestingly, we have found the new form of the solutions of $\phi(t)=\sqrt{c_{1}+c_{2}t+c_{3}e^{-\lambda t}}$ for GR and $\phi(t)=c_{7}^{-1/2}\sqrt{ c_{4}+c_{5}e^{-\lambda_{1}t}+c_{6}e^{{-\lambda_{2}t}}}$ for NMDC universe, respectively. However, it is very interesting to figure out the physical consequences of the found solutions and we leave this for further investigation. Finally, it is hoped that this work will stimulate for further research on seeking other possible solutions of the system equations based on Noether gauge symmetry as shown in Eq.(\ref{alpha a1})-Eq.(\ref{Bt_gaugeterm}) for GR  
 and  Eq.(\ref{cs0})-Eq.(\ref{Bt term}) for NMDC cosmology. 


\subsection*{Acknowledgments}
P. Channuie acknowledged the Mid-Career Research Grant 2020 from National Research Council of Thailand under a contract No. NFS6400117.
M. Dolohtahe acknowledged the Faculty of Science Research Fund, Prince of Songkla University, under a contract No.1-2561-02-011


\begin{thebibliography}{99}

\bibitem{SupernovaCosmologyProject:1998vns}
S.~Perlmutter \textit{et al.} [Supernova Cosmology Project],
Astrophys. J. \textbf{517} (1999), 565-586
doi:10.1086/307221
[arXiv:astro-ph/9812133 [astro-ph]].

\bibitem{SupernovaSearchTeam:1998fmf}
A.~G.~Riess \textit{et al.} [Supernova Search Team],
Astron. J. \textbf{116} (1998), 1009-1038
doi:10.1086/300499
[arXiv:astro-ph/9805201 [astro-ph]].

\bibitem{BICEP2:2015xme}
P.~A.~R.~Ade \textit{et al.} [BICEP2 and Keck Array],
Phys. Rev. Lett. \textbf{116} (2016), 031302
doi:10.1103/PhysRevLett.116.031302
[arXiv:1510.09217 [astro-ph.CO]].

\bibitem{WMAP:2010qai}
E.~Komatsu \textit{et al.} [WMAP],
Astrophys. J. Suppl. \textbf{192} (2011), 18
doi:10.1088/0067-0049/192/2/18
[arXiv:1001.4538 [astro-ph.CO]].

\bibitem{WMAP:2012nax}
G.~Hinshaw \textit{et al.} [WMAP],
Astrophys. J. Suppl. \textbf{208} (2013), 19
doi:10.1088/0067-0049/208/2/19
[arXiv:1212.5226 [astro-ph.CO]].

\bibitem{Planck:2015fie}
P.~A.~R.~Ade \textit{et al.} [Planck],
Astron. Astrophys. \textbf{594} (2016), A13
doi:10.1051/0004-6361/201525830
[arXiv:1502.01589 [astro-ph.CO]].

\bibitem{Planck:2015sxf}
P.~A.~R.~Ade \textit{et al.} [Planck],
Astron. Astrophys. \textbf{594} (2016), A20
doi:10.1051/0004-6361/201525898
[arXiv:1502.02114 [astro-ph.CO]].

\bibitem{BICEP2:2014owc}
P.~A.~R.~Ade \textit{et al.} [BICEP2],
Phys. Rev. Lett. \textbf{112} (2014) no.24, 241101
doi:10.1103/PhysRevLett.112.241101
[arXiv:1403.3985 [astro-ph.CO]].

\bibitem{BICEP2:2015nss}
P.~A.~R.~Ade \textit{et al.} [BICEP2 and Planck],
Phys. Rev. Lett. \textbf{114} (2015), 101301
doi:10.1103/PhysRevLett.114.101301
[arXiv:1502.00612 [astro-ph.CO]].

\bibitem{SDSS:2003eyi}
M.~Tegmark \textit{et al.} [SDSS],
Phys. Rev. D \textbf{69} (2004), 103501
doi:10.1103/PhysRevD.69.103501
[arXiv:astro-ph/0310723 [astro-ph]].

\bibitem{SDSS:2004kqt}
U.~Seljak \textit{et al.} [SDSS],
Phys. Rev. D \textbf{71} (2005), 103515
doi:10.1103/PhysRevD.71.103515
[arXiv:astro-ph/0407372 [astro-ph]].

\bibitem{SDSS:2005xqv}
D.~J.~Eisenstein \textit{et al.} [SDSS],
Astrophys. J. \textbf{633} (2005), 560-574
doi:10.1086/466512
[arXiv:astro-ph/0501171 [astro-ph]].

\bibitem{Planck:2018nkj}
N.~Aghanim \textit{et al.} [Planck],
Astron. Astrophys. \textbf{641} (2020), A1
doi:10.1051/0004-6361/201833880
[arXiv:1807.06205 [astro-ph.CO]].

\bibitem{Weinberg:2000yb}
S.~Weinberg,
[arXiv:astro-ph/0005265 [astro-ph]].

\bibitem{Velten:2014nra}
H.~E.~S.~Velten, R.~F.~vom Marttens and W.~Zimdahl,
Eur. Phys. J. C \textbf{74} (2014) no.11, 3160
doi:10.1140/epjc/s10052-014-3160-4
[arXiv:1410.2509 [astro-ph.CO]].

\bibitem{Nojiri:2010wj}
S.~Nojiri and S.~D.~Odintsov,
Phys. Rept. \textbf{505} (2011), 59-144
doi:10.1016/j.physrep.2011.04.001
[arXiv:1011.0544 [gr-qc]].

\bibitem{Nojiri:2006ri}
S.~Nojiri and S.~D.~Odintsov,
eConf \textbf{C0602061} (2006), 06
doi:10.1142/S0219887807001928
[arXiv:hep-th/0601213 [hep-th]].

\bibitem{Capozziello:2011et}
S.~Capozziello and M.~De Laurentis,
Phys. Rept. \textbf{509} (2011), 167-321
doi:10.1016/j.physrep.2011.09.003
[arXiv:1108.6266 [gr-qc]].

\bibitem{Bamba:2015uma}
K.~Bamba and S.~D.~Odintsov,
Symmetry \textbf{7} (2015) no.1, 220-240
doi:10.3390/sym7010220
[arXiv:1503.00442 [hep-th]].

\bibitem{Copeland:2006wr}
E.~J.~Copeland, M.~Sami and S.~Tsujikawa,
Int. J. Mod. Phys. D \textbf{15} (2006), 1753-1936
doi:10.1142/S021827180600942X
[arXiv:hep-th/0603057 [hep-th]].

\bibitem{Bergmann:1968ve}
P.~G.~Bergmann,
Int. J. Theor. Phys. \textbf{1} (1968), 25-36
doi:10.1007/BF00668828

\bibitem{Buchdahl:1970ynr}
H.~A.~Buchdahl,
Mon. Not. Roy. Astron. Soc. \textbf{150} (1970), 1

\bibitem{Sotiriou:2008rp}
T.~P.~Sotiriou and V.~Faraoni,
Rev. Mod. Phys. \textbf{82} (2010), 451-497
doi:10.1103/RevModPhys.82.451
[arXiv:0805.1726 [gr-qc]].

\bibitem{DeFelice:2010aj}
A.~De Felice and S.~Tsujikawa,
Living Rev. Rel. \textbf{13} (2010), 3
doi:10.12942/lrr-2010-3
[arXiv:1002.4928 [gr-qc]].

\bibitem{BeltranJimenez:2017doy}
J.~Beltran Jimenez, L.~Heisenberg, G.~J.~Olmo and D.~Rubiera-Garcia,
Phys. Rept. \textbf{727} (2018), 1-129
doi:10.1016/j.physrep.2017.11.001
[arXiv:1704.03351 [gr-qc]].

\bibitem{Nojiri:2017ncd}
S.~Nojiri, S.~D.~Odintsov and V.~K.~Oikonomou,
Phys. Rept. \textbf{692} (2017), 1-104
doi:10.1016/j.physrep.2017.06.001
[arXiv:1705.11098 [gr-qc]].


\bibitem{Amendola:1993uh}
L.~Amendola,
Phys. Lett. B \textbf{301} (1993), 175-182
doi:10.1016/0370-2693(93)90685-B
[arXiv:gr-qc/9302010 [gr-qc]].

\bibitem{Capozziello:1999xt}
S.~Capozziello, G.~Lambiase and H.~J.~Schmidt,
Annalen Phys. \textbf{9} (2000), 39-48
doi:10.1002/(SICI)1521-3889(200001)9:1\ensuremath{<}39::AID-ANDP39\ensuremath{>}3.0.CO
[arXiv:gr-qc/9906051 [gr-qc]].

\bibitem{Capozziello:1999uwa}
S.~Capozziello and G.~Lambiase,
Gen. Rel. Grav. \textbf{31} (1999), 1005-1014
doi:10.1023/A:1026631531309
[arXiv:gr-qc/9901051 [gr-qc]].

\bibitem{Granda:2010hb}
L.~N.~Granda and W.~Cardona,
JCAP \textbf{07} (2010), 021
doi:10.1088/1475-7516/2010/07/021
[arXiv:1005.2716 [hep-th]].

\bibitem{Granda:2010ex}
L.~N.~Granda,
Class. Quant. Grav. \textbf{28} (2011), 025006
doi:10.1088/0264-9381/28/2/025006
[arXiv:1009.3964 [hep-th]].


\bibitem{Sushkov:2009hk}
S.~V.~Sushkov,
Phys. Rev. D \textbf{80} (2009), 103505
doi:10.1103/PhysRevD.80.103505
[arXiv:0910.0980 [gr-qc]].

\bibitem{Saridakis:2010mf}
E.~N.~Saridakis and S.~V.~Sushkov,
Phys. Rev. D \textbf{81} (2010), 083510
doi:10.1103/PhysRevD.81.083510
[arXiv:1002.3478 [gr-qc]].

\bibitem{Skugoreva:2013ooa}
M.~A.~Skugoreva, S.~V.~Sushkov and A.~V.~Toporensky,
Phys. Rev. D \textbf{88} (2013), 083539
[erratum: Phys. Rev. D \textbf{88} (2013) no.10, 109906]
doi:10.1103/PhysRevD.88.083539
[arXiv:1306.5090 [gr-qc]].

\bibitem{Sushkov:2012za}
S.~Sushkov,
Phys. Rev. D \textbf{85} (2012), 123520
doi:10.1103/PhysRevD.85.123520
[arXiv:1204.6372 [gr-qc]].

\bibitem{Koutsoumbas:2013boa}
G.~Koutsoumbas, K.~Ntrekis and E.~Papantonopoulos,
JCAP \textbf{08} (2013), 027
doi:10.1088/1475-7516/2013/08/027
[arXiv:1305.5741 [gr-qc]].

\bibitem{Gumjudpai:2015vio}
B.~Gumjudpai and P.~Rangdee,
Gen. Rel. Grav. \textbf{47} (2015) no.11, 140
doi:10.1007/s10714-015-1985-2
[arXiv:1511.00491 [gr-qc]].

\bibitem{Gumjudpai:2016frh}
B.~Gumjudpai, Y.~Jawralee and N.~Kaewkhao,
Gen. Rel. Grav. \textbf{49} (2017) no.9, 120
doi:10.1007/s10714-017-2287-7
[arXiv:1609.08189 [gr-qc]].


\bibitem{Granda:2011zk}
L.~N.~Granda,
JCAP \textbf{04} (2011), 016
doi:10.1088/1475-7516/2011/04/016
[arXiv:1104.2253 [hep-th]].

\bibitem{Germani:2010gm}
C.~Germani and A.~Kehagias,
Phys. Rev. Lett. \textbf{105} (2010), 011302
doi:10.1103/PhysRevLett.105.011302
[arXiv:1003.2635 [hep-ph]].

\bibitem{Sadjadi:2012zp}
H.~M.~Sadjadi and P.~Goodarzi,
JCAP \textbf{02} (2013), 038
doi:10.1088/1475-7516/2013/02/038
[arXiv:1203.1580 [gr-qc]].

\bibitem{Tsujikawa:2012mk}
S.~Tsujikawa,
Phys. Rev. D \textbf{85} (2012), 083518
doi:10.1103/PhysRevD.85.083518
[arXiv:1201.5926 [astro-ph.CO]].

\bibitem{Ema:2015oaa}
Y.~Ema, R.~Jinno, K.~Mukaida and K.~Nakayama,
JCAP \textbf{10} (2015), 020
doi:10.1088/1475-7516/2015/10/020
[arXiv:1504.07119 [gr-qc]].

\bibitem{Yang:2015pga}
N.~Yang, Q.~Fei, Q.~Gao and Y.~Gong,
Class. Quant. Grav. \textbf{33} (2016) no.20, 205001
doi:10.1088/0264-9381/33/20/205001
[arXiv:1504.05839 [gr-qc]].

\bibitem{Myung:2015tga}
Y.~S.~Myung, T.~Moon and B.~H.~Lee,
JCAP \textbf{10} (2015), 007
doi:10.1088/1475-7516/2015/10/007
[arXiv:1505.04027 [gr-qc]].

\bibitem{MohseniSadjadi:2013iou}
H.~Mohseni Sadjadi and P.~Goodarzi,
Phys. Lett. B \textbf{732} (2014), 278-284
doi:10.1016/j.physletb.2014.03.050
[arXiv:1309.2932 [astro-ph.CO]].

\bibitem{Capozziello:1994dn}
S.~Capozziello, R.~De Ritis and P.~Scudellaro,
Nuovo Cim. B \textbf{109} (1994), 159-165
doi:10.1007/BF02727426

\bibitem{Channuie:2017txg}
P.~Channuie and D.~Momeni,
Nucl. Phys. B \textbf{935} (2018), 256-270
doi:10.1016/j.nuclphysb.2018.08.016
[arXiv:1712.07927 [gr-qc]].

\bibitem{Nurbaki:2020dgw}
A.~N.~Nurbaki, S.~Capozziello and C.~Deliduman,
Eur. Phys. J. C \textbf{80} (2020) no.2, 108
doi:10.1140/epjc/s10052-020-7666-7
[arXiv:2001.02304 [gr-qc]].

\bibitem{Jamil:2012fs}
M.~Jamil, D.~Momeni and R.~Myrzakulov,
Eur. Phys. J. C \textbf{72} (2012), 2137
doi:10.1140/epjc/s10052-012-2137-4
[arXiv:1210.0001 [physics.gen-ph]].

\bibitem{Momeni:2015gka}
D.~Momeni, R.~Myrzakulov and E.~G\"udekli,
Int. J. Geom. Meth. Mod. Phys. \textbf{12} (2015) no.10, 1550101
doi:10.1142/S0219887815501017
[arXiv:1502.00977 [gr-qc]].

\bibitem{Capozziello:2008ch}
S.~Capozziello and A.~De Felice,
JCAP \textbf{08} (2008), 016
doi:10.1088/1475-7516/2008/08/016
[arXiv:0804.2163 [gr-qc]].

\bibitem{Kaewkhao:2017evn}
N.~Kaewkhao, T.~Kanesom and P.~Channuie,
Nucl. Phys. B \textbf{931} (2018), 216-225
doi:10.1016/j.nuclphysb.2018.04.011
[arXiv:1711.10080 [gr-qc]].

\bibitem{Bajardi:2020osh}
F.~Bajardi and S.~Capozziello,
Eur. Phys. J. C \textbf{80} (2020) no.8, 704
doi:10.1140/epjc/s10052-020-8258-2
[arXiv:2005.08313 [gr-qc]].

\bibitem{Camci:2007zz}
U.~Camci and Y.~Kucukakca,
Phys. Rev. D \textbf{76} (2007), 084023
doi:10.1103/PhysRevD.76.084023

\bibitem{Jamil:2012zm}
M.~Jamil, S.~Ali, D.~Momeni and R.~Myrzakulov,
Eur. Phys. J. C \textbf{72} (2012), 1998
doi:10.1140/epjc/s10052-012-1998-x
[arXiv:1201.0895 [physics.gen-ph]].

\bibitem{Channuie:2018now}
P.~Channuie, D.~Momeni and M.~A.~Ajmi,
Eur. Phys. J. C \textbf{78} (2018) no.7, 588
doi:10.1140/epjc/s10052-018-6061-0
[arXiv:1807.07406 [physics.gen-ph]].

\bibitem{Oz:2021nmj}
I.~B.~\"Oz and K.~Bamba,
[arXiv:2109.08938 [gr-qc]].


\bibitem{Aslam:2013pga}
A.~Aslam, M.~Jamil, D.~Momeni, R.~Myrzakulov, M.~A.~Rashid and M.~Raza,
Astrophys. Space Sci. \textbf{348} (2013), 533-540
doi:10.1007/s10509-013-1569-0
[arXiv:1308.2221 [astro-ph.CO]].

\bibitem{Hussain:2011wa}
I.~Hussain, M.~Jamil and F.~M.~Mahomed,
Astrophys. Space Sci. \textbf{337} (2012), 373-377
doi:10.1007/s10509-011-0812-9
[arXiv:1107.5211 [physics.gen-ph]].

\bibitem{Kucuakca:2011np}
Y.~Kucuakca and U.~Camci,
Astrophys. Space Sci. \textbf{338} (2012), 211-216
doi:10.1007/s10509-011-0921-5
[arXiv:1111.5336 [gr-qc]].

\bibitem{Kanesom:2021ytb}
T.~Kanesom, P.~Channuie and N.~Kaewkhao,
Eur. Phys. J. C \textbf{81} (2021) no.4, 357
doi:10.1140/epjc/s10052-021-09144-2
[arXiv:2102.07633 [gr-qc]].

\bibitem{Henneaux:1992ig}
Henneaux M and Teitelboim C 1992
\textit{Quantization of gauge systems} (
Princeton: Princeton University Press) pp 4-5

\bibitem{Avdeev:2021von}
N.~Avdeev and A.~Toporensky,
[arXiv:2103.00556 [gr-qc]].


\bibitem{Babara Ryden 2017}
B. Ryden, 2017 \textit{Introduction to Cosmology $2^{nd}$ Edition} (New York: Cambridge University Press)

\bibitem{Bouhmadi-Lopez:2016dcf}
Bouhmadi-López M and Chen C 2016
\textit{JCAP }\textbf{11} 023 (\textit{Preprint} \rm gr-qc/1609.00700)

\bibitem{SARLET-1981}
Sarlet W, Kntrijin R, Generalizations of Noether’s theorem in classical mechanics. SIAM Rev. 28, 467–484 (1981)

\bibitem{Leone R-2015}
Leone R and Gourieux T, Eur. J. Phys, vol. 36 (2015), p. 065022

\bibitem{Mukherjee:2021bna}
A.~Mukherjee and S.~B.~Roy,
[arXiv:2102.10483 [gr-qc]].

\bibitem{Handley:2014bqa}
W.~J.~Handley, S.~D.~Brechet, A.~N.~Lasenby and M.~P.~Hobson,
Phys. Rev. D \textbf{89}, no.6, 063505 (2014)
doi:10.1103/PhysRevD.89.063505
[arXiv:1401.2253 [astro-ph.CO]].

\bibitem{Harko:2016xip}
T.~Harko, F.~S.~N.~Lobo, E.~N.~Saridakis and M.~Tsoukalas,
Phys. Rev. D \textbf{95}, no.4, 044019 (2017)
doi:10.1103/PhysRevD.95.044019
[arXiv:1609.01503 [gr-qc]].

\bibitem{Darabi:2013caa}
F.~Darabi and A.~Parsiya,
Class. Quant. Grav. \textbf{32}, no.15, 155005 (2015)
doi:10.1088/0264-9381/32/15/155005
[arXiv:1312.1322 [gr-qc]].


\end{thebibliography}
\end{document}